\documentclass[twocolumn]{svjour3}          
\smartqed  
\RequirePackage{fix-cm}

\usepackage[dvipdfmx]{graphicx} \usepackage{bmpsize}

\usepackage{epsfig}
\usepackage{dcolumn}
\usepackage{bm}
\usepackage{amsmath}
\usepackage{amsfonts}
\usepackage{amssymb}
\usepackage{graphicx}
\usepackage{latexsym}
\usepackage{rotating}
\usepackage{multirow}
\usepackage{slashed}
\usepackage{color}
\usepackage{hyperref}

\usepackage[shortlabels]{enumitem}
\setlist[enumerate, 1]{1\textsuperscript{o}}

\begin{document}

\title{Recursive Neural Networks in Quark/Gluon Tagging}

\author{Taoli Cheng}

\institute{Taoli Cheng \at
School of Physical Sciences,\\
University of Chinese Academy of Sciences, 100049 Beijing,\\
P. R. China\\
        \email{chengtaoli.1990@gmail.com} 
}


\date{\today}

\maketitle

\begin{abstract}
Since the machine learning techniques are improving rapidly, it has been shown that
the image recognition techniques in deep neural networks can be used to detect jet substructure. And it turns 
out that 
deep neural networks can match or outperform traditional approach of 
expert features. However, there are disadvantages such as sparseness of jet images.
Based on the natural tree-like structure of jet 
sequential clustering, the recursive neural networks (RecNNs), which 
embed jet clustering history recursively as in natural language processing, 
have a better behavior when confronted with these problems.
We thus try to explore the performance of RecNNs
in quark/gluon discrimination.
The results show that RecNNs work better than the baseline boosted decision tree (BDT) by a few percent in 
gluon rejection rate. However, extra implementation
of particle flow identification only increases the performance slightly.
We also experimented on some relevant aspects which might influence the 
performance of the networks. It shows that even taking only particle flow identification as input feature
without any extra information on momentum or angular position
is already giving a fairly good result, which indicates that the most of the information
for quark/gluon discrimination is already included in the tree-structure itself.
As a bonus, a rough up/down quark jets discrimination is also explored.
\keywords{Jet Physics \and LHC Phenomenology \and Deep Learning}
\end{abstract}

\section{Introduction}

The study of jet substructure has been the very advanced topic in jet physics 
at the Large Hadron Collider (LHC). 
The techniques in distinguishing different substructures have been established well \cite{Larkoski:2017jix,Adams:2015hiv}.
And at this stage,  different jet observables have been invented to improve our understanding about jets. 
As for application, we can use jet substructure to 
detect new physics such as Supersymmetry or Two Higgs Doublet Models. 
For theoretical development, demand on 
theoretical calculation of jet substructure observables actually pushes our 
precision calculation forward and helps us understand QCD 
better. And the development of Monte Carlo tools also interacts with jet measurements
at the LHC.

When the experimental environment gets more complex and the problem has larger and larger dimension, 
the number of observables we need is increasing, and sometimes gets so large that exceeds our actual computation capability
. Artificial neural networks have already been 
employed for high dimensional problems. Earlier research using neural networks are carried out in a framework of 
designing observables by hand at first and then feeding these observables into neural 
networks to do classification. It strongly relies on the pre-stage of 
expert-features designing, and thus depends on physicists' understanding on the 
problem. The expert-feature approach has a very long history in high energy physics, 
but there are several pros and cons.
They generally have clear physical intuition. The observables are designed according to 
physical understanding and theoretical insights. And their behavior are 
well understood, and based on theoretical framework.
However, they can only deal with problems case by case,
and highly depend on the specific processes worked with. 
And there are always strong correlation between observables. 
At last, since we are not guaranteed that all the information 
can be captured in the observables, the best we can do is just approaching the limit by
trial and error. And there is no guide for how much information we have captured.

Along with the new framework in machine learning (ML) getting more and more mature, relevant
techniques have been employed in high energy physics. The new
machine learning framework augmented our ability to fully utilize experimental 
data. On one hand, the input data can be taken from the raw detector 
measurements, which means we don't have to lose information because of data 
transformation or specific observable designing. On the other hand, this new 
input formulation also brings new insights on how we organize our observations.
Finally, the uniform formulation and general approach might help us out of the
busy tasks of too-many-sub-channels designing.
Especially for jets and their structure analysis, we have the opportunity to improve our 
working culture to adapt to the ML era.

There have already been some works using deep neural networks (DNN) in jet physics. The very first 
attempt was made in using computer vision \cite{Cogan:2014oua} to help with jet tagging.
And later people started to use image recognition in boosted top 
tagging \cite{Almeida:2015jua,Pearkes:2017hku,Kasieczka:2017nvn}, boosted W 
tagging \cite{Baldi:2016fql}, heavy flavour classification \cite{Guest:2016iqz}, 
and the investigation of parton shower uncertainties is also 
made \cite{Barnard:2016qma}. A detailed report on image recognition in jet 
substructure is given in reference \cite{deOliveira:2015xxd}. And more 
interestingly, colored version of image recognition inspired by particle flow has
also been proposed \cite{Komiske:2016rsd}. Based on all these recent 
development, it can be shown that deep neural networks generally match or 
outperform the conventional expert feature approach. And the colored version 
can perform better than gray scale in some cases since it employs more 
information. The basic idea of image recognition in jet physics is mapping a jet onto 
the $({\rm azimuthal~angle}~ \eta, ~{\rm pseudorapidity}~ \phi)$ plane and translating the transverse momentum $p_T$ of every constituent of the 
jet into intensity of the pixel. Thus the higher $p_T$, the darker the pixel 
will 
be. By feeding these jet images after some preprocessing into deep neural 
networks, we can discriminate signal from background with the raw data from the 
detector while avoiding too much human designing on the physics problem. 
And at experiments, CMS has already carried out heavy flavor identification 
using DNN recently \cite{CMS-DP-2017-005}.

Until now, most of the work is done in the regime of image recognition, since 
it is the most intuitive, simple and general approach. However, there are some 
disadvantages for image recognition:
\begin{itemize}
 \item sparseness: in most cases only 5\%-10\% pixels are 
active in jet images of fixed size \cite{deOliveira:2015xxd}. Thus most of the
parameter space is actually wasted.
 \item the pixelization process causes information 
loss. 
 \item computation cost: too many model parameters require more computation power. 
 Not very efficient for larger image size.
 \item it's more complicated to do the event-level analysis. 
Of course one can channel the output of image classifier into later classifier based on 
 expert features or other network architecture, but that will anyway bring
 some complication and detailed investigation is necessary.
\end{itemize}

Aside from image recognition, another natural approach is recurrent neural networks (RNN)
\cite{Goodfellow-et-al-2016}, which takes sequences as input and
is widely used in natural language processing.
RNN adapts to the problems confronted in image recognition 
better since it can properly deal with sequential input of variable length.
And also the parameter sharing makes it very efficient.
To be more concrete, a recurrent network is 
structured as a recurrent chain, in which every step takes in the output of the 
last computation, and all the steps share the same set of parameters. Recursive 
Neural Networks (here abbreviated as RecNN in order not to be confused with recurrent 
neural networks), rather, has a tree-like structure, other than the chain-like 
one of RNN. Respect to RNN, RecNN reduces the computation depth from 
$\tau$ to $\mathcal{O}(\log \tau)$. 
As an example, RNN is explored in \cite{Guest:2016iqz} for heavy flavor tagging, by taking 
low level track and vertex information as input for the neural networks.

Base on the natural analogy of the sequentially clustered jet and the input structure of RecNN,
a group \cite{Louppe:2017ipp} has  implemented 
the RecNN version of jet analysis. In their work, the framework is built for 
embedding jet recursively into a root jet node, and then feeding it into the 
sequential classifier. And by a simple extension, event-level analysis can be 
easily implemented in a structure-in-structure manner, although right now only 
dealing with jets-only events with very limited application range.
The work is done in the regime of jet substructure for 
boosted gauge boson. And the results show that the RecNN outperforms the
expert feature approach and image version in DNN. 

Motivated by all these progresses and the prospects, we try to explore the 
performance of RecNN in another very interesting topic: quark/gluon 
discrimination. Quark/gluon tagging is gaining great potential at the LHC.
Since gluon has a larger color factor than quarks, 
gluon jets will generally have more radiation and also broader radiation pattern. The ratio of 
final state counts for gluon and quark jets is predicted by an asymptotic value of 9/4. 
The measurements at the LHC \cite{Aad:2016oit} also support this 
prediction, where the charged particle multiplicity is measured, and shows the 
tendency of approaching the limit.

\begin{figure*}[ht!]
 \begin{center}
  \includegraphics[width=0.3\linewidth]{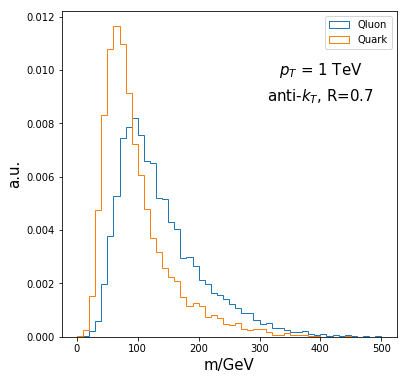}
  \includegraphics[width=0.3\linewidth]{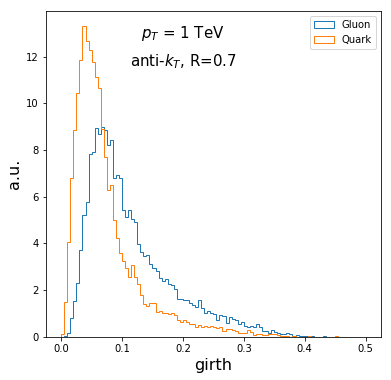}
  \includegraphics[width=0.3\linewidth]{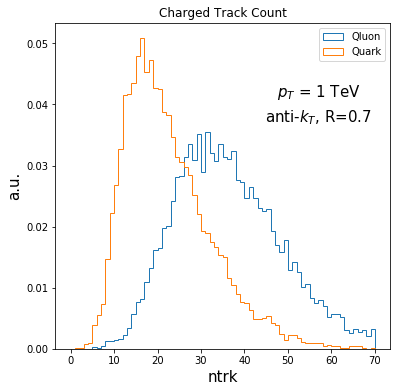}
  \caption{\label{jet_obs}Distributions of jet scaled mass, girth, and track count 
for quark and gluon jets with $p_T=$ 1 TeV.}
 \end{center}
\end{figure*}

The conventional approach in quark/gluon discrimination is 
defining some jet observables, such as charge multiplicity, jet mass, jet 
subjettiness \cite{Gallicchio:2012ez} (Fig. \ref{jet_obs} shows the 
distributions of a few observables for 1 TeV quark and gluon jets). 
These jet observables have turned to 
be very efficient and well-motivated discriminants. For a multivariable analysis, 
the general performance is: for a 50\% quark acceptance, the gluon rejection can 
reach 80 \% - 90 \%, and the corresponding significance can have an increase by a 
factor of 2 - 3.
These results show a great potential in helping new physics search, and thus the
exploration in DNN which can make use of all the low level information is worthwhile.
The image approach using Convolutional Neural Networks (CNN) has been explored \cite{Komiske:2016rsd}, showing 
performance matching or outperforming conventional expert feature approach. And
the DNN simulation from ATLAS \cite{ATL-PHYS-PUB-2017-017} and CMS \cite{CMS-DP-2017-027}
using either CNN or RNN also confirm the potential. For a 50\% quark acceptance at jet $p_T \sim 200$
GeV, mis-identification rate is approximately $10\%$ (a few percent away from the pythia-level results).

In this work, we aim to address the following questions:
\begin{itemize}
 \item At first, how does the RecNN generally perform in quark/gluon discrimination at the LHC.
 \item We include fast detector simulation in our analysis, trying to bring a more 
realistic picture to the problem.
 \item How to better implement particle flow identification in RecNN, in order to gain 
 a more containing information set.
\end{itemize}

The paper is organized as following: in Section \ref{sec:meth}, we describe in detail the 
process of jet embedding and the neural network architecture. Then we explore the
RecNN performance in quark/gluon discrimination in Section \ref{sec:rst}. 
Besides, a brief investigation in up/down quark discrimination using RecNN is also presented in 
Section \ref{sec:rst}.
Conclusions are given
in Section \ref{sec:concls}. And finally a short discussion and outlook in Section \ref{sec:disc} is presented.

\section{\label{sec:meth}Methodology}

\subsection{Jets at the LHC}

There are several different definitions \cite{Gras:2017jty} of a "jet".
How to connect the theoretical parton and the experimental measurements
on a collimated spray of hadrons is not a trivial problem.
But in practice, the jet is operationally defined by the 
jet algorithm used to group particles into clusters.

The basic idea of sequential jet clustering \cite{Salam:2009jx} is to recombine the particles into 
pseudojets recurrently according to some measure 
and recombination scheme. The ``closest" particles will be combined
into pseudojets at first. And by this QCD-inspired
measure, we can get soft and collinear safe jet definition.
And through this recursive clustering process, the global information 
including number of particles, their momenta, clustering history, is embedded in 
the tree structure of the clustering sequence.

\begin{figure}[ht!]
 \begin{center}
  \includegraphics[width=0.25\textwidth,angle=180]{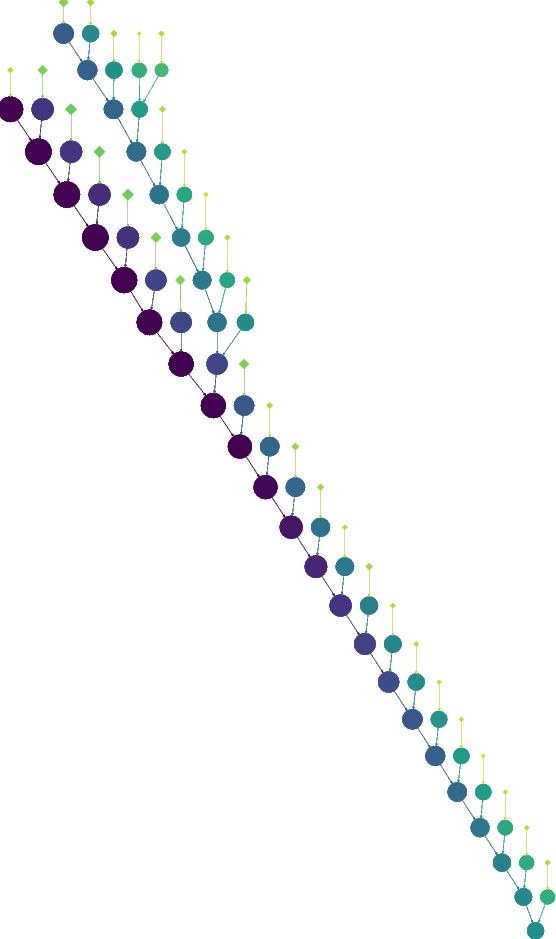}
  \includegraphics[width=0.35\textwidth,angle=180]{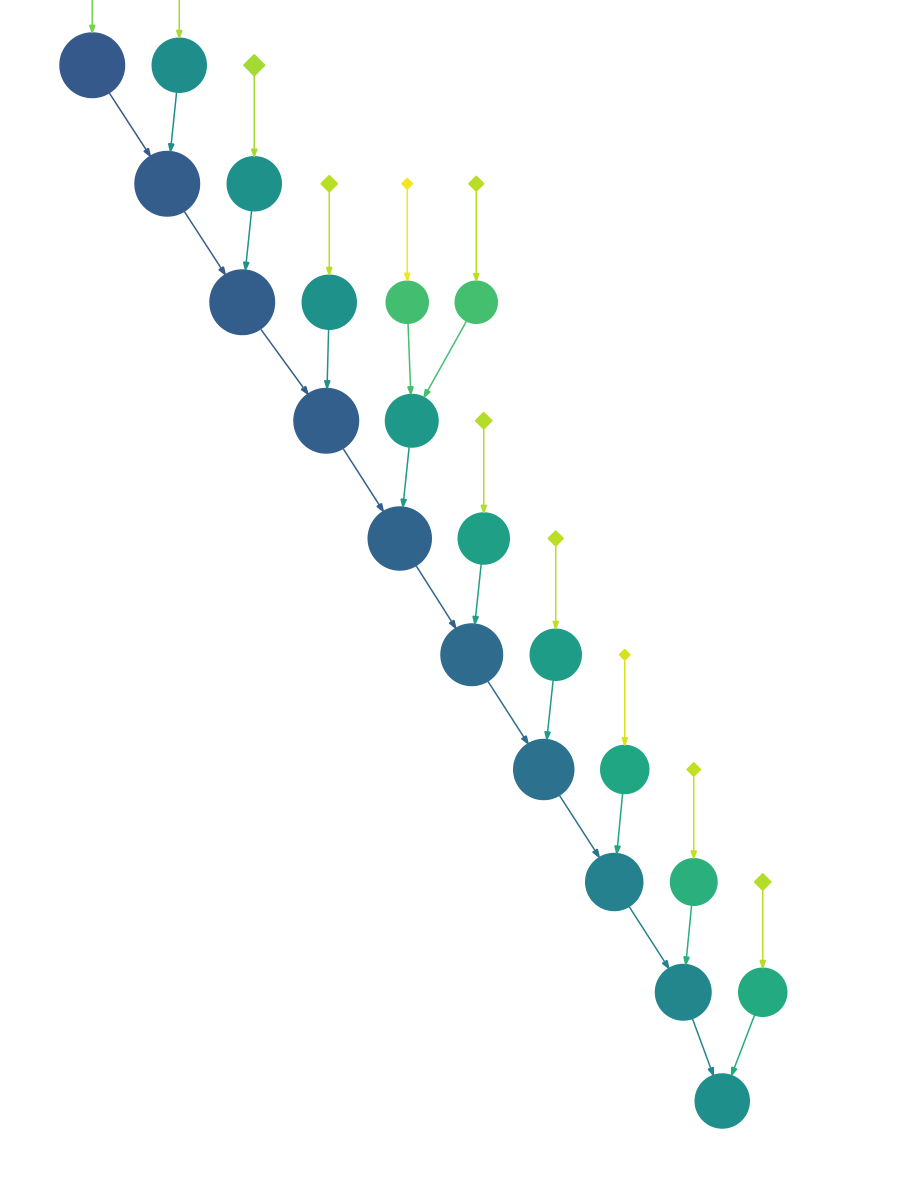}
  \caption{\label{trees}Typical tree structures for 1 TeV gluon jet (left) and quark jet (right).}
 \end{center}
\end{figure}

Here shows the typical tree-structures of a 1 TeV gluon jet and quark jet in Fig. \ref{trees}.

A naive impression is that the gluon jet has more constituents and accordingly more
complicated clustering structure.
Actually the mean content number of a 1 TeV gluon jet is $\sim$ 90, while for a quark jet is 
approximately 50. And the typical number of charged constituents for a 1 TeV 
gluon jet is 36, and for a quark jet is 22. 
Thus if we input all the measured degrees of freedom (d.o.f.) into analysis, for every 
final state particle we have 3 d.o.f. ($p_T$, $\eta$ and $\phi$) and one more 
if particle flow identification is taken into account, thus for a whole jet 
the total d.o.f can reach $\sim 4 \times 100$. 
If we want to utilize all the information we've gotten, such high dimensionality 
naturally brings us to the DNN approach, although
there are arguments about how much information actually is included in a 
jet \cite{Datta:2017rhs}. 

\subsection{Network Architecture}

\begin{figure}
 \begin{center}
  \includegraphics[width=0.5\textwidth]{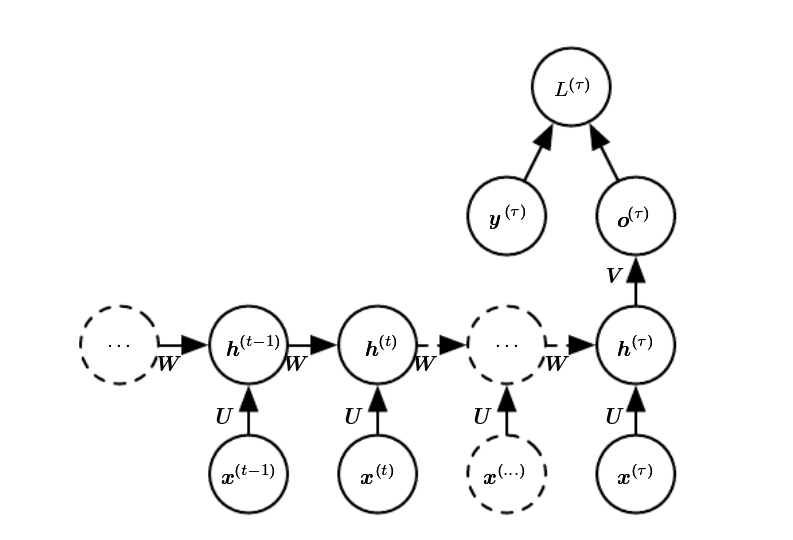}
  \includegraphics[width=0.45\textwidth]{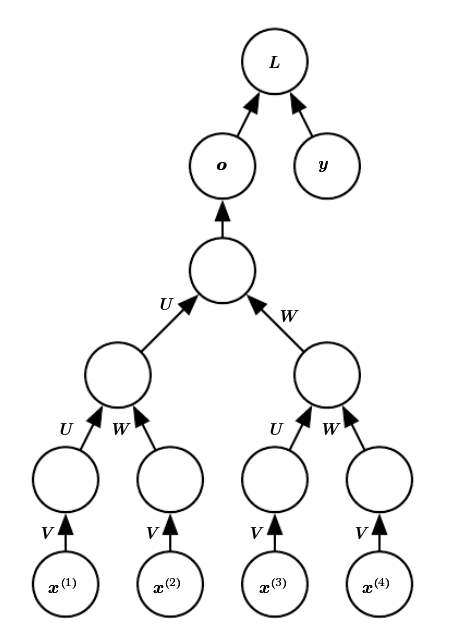}
  \caption{\label{rnn} \textbf{Left}: Time-unfolded computational graph for a RNN which only 
takes the output at the end of the sequence. 
\textbf{Right}: Time-unfolded computational graph for a RecNN. Plots are taken from Ref. 
\cite{Goodfellow-et-al-2016} with permission.}
 \end{center}
\end{figure}

Here first gives a short introduction to RNN.
Recurrent neural networks are initially designed for processing sequential 
data. The state of the system is recurrently defined. For the t-th step, the 
state ${\bf h}^{(t)}$ is defined by the output of last step and the new feed as:
\begin{eqnarray}\notag
 \textbf{h}^{(t)} &=& g^{(t)} (x^{(t)}, x^{(t-1)},..., x^{(1)})  \\ 
         &=& f(h^{(t-1)}, x^{(t)}; \theta)
\end{eqnarray}
where the same transition function f is applied for every step. Thus
this parameter sharing makes the model very simple and efficient.
And the recursive networks, rather, use a tree-structure instead of a chain-like 
structure. As shown in the right part of Fig. \ref{rnn}, the matrices {\bf W}, {\bf U}
and {\bf V} are shared parameters by all the steps.
In Fig. \ref{rnn}, unfolded computational graphs of a typical RNN and a RecNN are depicted.
And there is one hidden-to-output at the end of the sequence. This output can be a final loss 
function or can be fed into following classifier. In brief, it summarizes a sequence and 
gives a fixed-length representation.

Based on the similarity between jet clustering and RecNN architecture, the application of RecNN
is straightforward.
The raw data from detector can be used as direct input to the networks. The 
measured transverse momentum $p_T$ and the angular location ($\eta,~\phi$) 
give the basic feature set.
After a regular jet clustering process, the jet (defined by the clustering 
structure {\bf t} with its contents $\{v_i, i = 1,..,N_j\}$, where $v_i$ denotes the 
four-momentum vector of i-th particle within the jet)
is embedded recursively into an embedding space of fixed size,
then the embedded jet node will be channeled to following classifier (a MLP).

According to Ref. \cite{Louppe:2017ipp}, a jet is recursively embedded 
into a single jet node ${\bf h}^{\rm jet}_1$. And through this recursive embedding, 
the history of jet clustering can be included in the final jet node. The 
procedure of embedding is a mapping from the feature space (with dimension f) 
to the embedding space (with dimension q) $R^f \to R^q$. For 
every jet with $N_j$ constituents, there are $2 N_j-1$ clustering nodes. Every 
node is represented by an input vector in the embedding space ${\bf u} \in  R^q$ 
after a transformation from feature vector {\bf x} (generally can be defined as 
($p_T$, $\eta$, $\phi$) or other version with the same information contents) as following:
\begin{equation}
\textbf{u}_k = \sigma(W_u {\bf x}_k + b_u)~\, \textrm{ for~the~k-th~node~~}
\end{equation}
where $W_u \in R^{q \times f}$, $b_u \in R^q$ and $\sigma$ denotes the ReLU 
activation function.

Then, the embedding of every node is defined by its children and 
its own input feed (we can also define the embedding by simply using the 
children. This version is also explored and shows similar performance while 
reducing the model parameters, see discussion in the next section): 
\begin{equation}
 \textbf{h}_k = \sigma\left( W_h  \left[
    \begin{array}{c}
    \textbf{h}^{\rm jet}_{k_L} \\ \textbf{h}^{\rm jet}_{k_R} \\ \textbf{u}_k
    \end{array} \right]
    + b_h
  \right) \label{eqn_embedding}    
\end{equation}
where we have model parameters $W_h \in R^{q \times 3q}$, $b_h \in R^q$. $k_L$
and $k_R$ are left and right child of node k respectively.
Down to the leaves (the original jet constituents), we have directly ${\bf h}_k 
= {\bf u}_k$. Thus, combining the raw input of detector measurements and jet 
clustering history, we can embed the information finally into a single root jet 
node ${\bf h}_1$, which is passed to a following classifier for final 
classification.
All the embedding parameters ($W_u$, $b_u$, $W_h$ and $b_h$) are learned using 
backpropagation jointly with the parameters of the following classifier $W_{\rm clf} {\rm ~and~} b_{\rm clf}$,
by trying to minimize the loss function.

The procedure is depicted as following:
\begin{equation}
 [\{{\bf t}, \{v_i, i=1,.., N_j\}\} \to {\bf h}^{\rm jet} \in R^q] \to {\rm ReLU} \to {\rm ReLU} \to {\rm Sigmoid}~\,
\end{equation}
where the rectified linear unit \cite{pmlr-v15-glorot11a} (ReLU) $= {\rm max}\{0,z\}$ is used for the hidden layers
in the classifier,
and Sigmoid ($\frac{1}{1+e^{-z}}$) activation is used for the output layer. And 
the log loss function is employed:
\begin{equation}
 L= - \frac{1}{N} \sum_i^N(y_i \log(y_i^{pred}) + (1-y_i) \log(1 - 
y_i^{pred}))~\,
\end{equation}
where $y_i$ is the label for i-th jet, and $y_i^{pred}$ is the prediction of the model.

\noindent\textbf{Recursively defined jet charge:}

Other than the very basic input set ($p_T$, $\eta$, $\phi$), more information 
can be included. At the LHC, the particle flow algorithm \cite{Sirunyan:2017ulk} combines the 
information through 
different parts of the detector thus gives more identification ability. 
It can match tracks to the energy deposit in the calorimeters, thus we 
have more accurate knowledge about the final states, and also higher precision 
on their transverse momenta.
Right 
now, we can identify charged tracks, neutral particles and photons within one 
jet. But how to implement this information in RecNN deserves some 
exploration. They can be implemented in a naive way as one-hot vectors 
(($i_{\rm neutral~hadron}$, $i_{\rm photon}$, $i_{\rm +}$, $i_{\rm -}$), $i=0~{\rm or}~1$, 
activated by the particle flow identification, e.g., a photon is represented by (0, 1, 0, 0))
added to the feature vector.
However, since the one-hot implementation doesn't 
have an additive nature (or, the particle flow identification of the inner nodes
is not well-defined), the recursive embedding can't utilize this information 
effectively. In Ref. \cite{Louppe:2017ipp}, the authors claimed that including particle flow
identification wouldn't gain any significant improvement for their attempt in 
discriminating boosted W jets and QCD jets.

In order to search for a better way to implement the particle flow identification,
we ask help from a jet observable: jet electric charge.
Jet charge is a very useful observable for identifying jet flavor and the identification of
W' and Z'. The pt-weighted jet charge 
\cite{Field:1977fa,Krohn:2012fg} is defined as following:
\begin{equation}
 Q_{\kappa}^J = \sum_{i \in J} (\frac{p_T^i}{p_T^J})^\kappa q_i ~, \label{ptwcharge}
\end{equation}
where $q_i$ is the electric charge of the particle within the jet, and $p_T^i$ 
is the transverse momentum of the component, while $p_T^J$ denotes the total 
transverse momentum of the jet. $\kappa$ is a free parameter, and $\kappa \to 
0$ gives the limit of simply adding charges of the components while $\kappa \to 
\infty$ gives the limit of the charge of the hardest component. The typical value
has been used by experiments lies between 0.2 and 1.0.

Trying to carry the particle flow information in the RecNN, we construct the 
recursively defined pt-weighted charge for the clustering tree:
\begin{equation}
 Q_{k}^{\rm rec} = \frac{Q_{k_L}^{\rm rec} (p_T^{k_L})^\kappa + Q_{k_R}^{\rm rec}
(p_T^{k_R})^\kappa}{(p_T^k)^\kappa}~\,
\end{equation}
while for the leaves $Q_k^{\rm rec}=q_i$.
The pt-weighted charge of k-th node $Q_{k}^{\rm rec}$ is defined by its children
in the same manner as in Eqn. \ref{ptwcharge}.
In this way, we still get the right pt-weighted 
charge for the jet node at the end of the embedding. This $Q_{k}^{\rm rec}$
along with the traditional feature set contributes to an almost-complete set of the 
information we get from detectors. $\kappa=0.5$ will be used throughout this work.

\begin{figure*}[ht!]
 \begin{center}
  \includegraphics[width=0.8\textwidth]{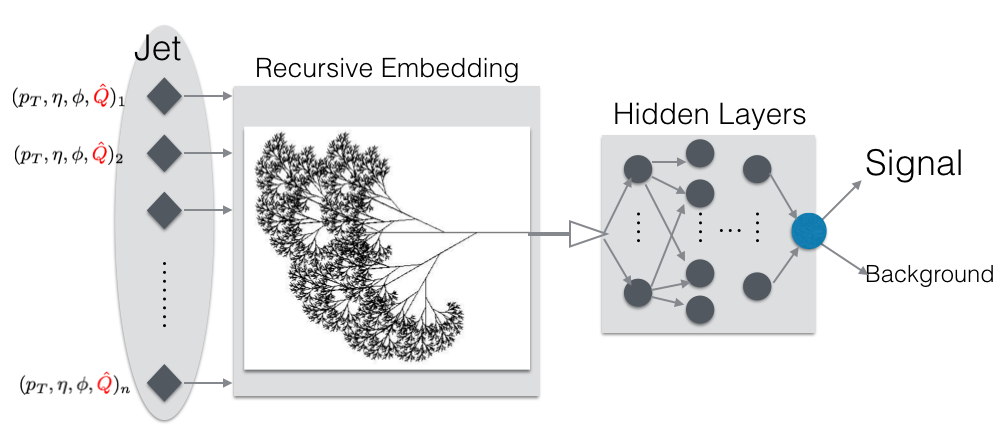}
  \caption{\label{jet-embedding} Workflow including jet embedding and the following classification (schematic
  plot of the recursion is credited to \textit{Brentsmith101} under \textit{Wikimedia Commons}).}
 \end{center}
\end{figure*}

Prepared with all the basic settings, we depict the whole architecture in Fig. \ref{jet-embedding}. 

\subsection{Setup}

\textbf{HEP Data Preparation:}

In preparing sample jets, Pythia8 \cite{Sjostrand:2014zea} is used to generate 
events and carry out the parton shower and the following haronization. Delphes 
\cite{deFavereau:2013fsa} is used for fast
detector simulation (pile-up effects are left aside for the time being). Jets are
first clustered using FastJet 
\cite{Cacciari:2011ma} with anti-kt algorithm \cite{Cacciari:2008gp}. 
After preprocessing,
the constituents within one jet will be reclustered and then 
the four-momenta of the particles in a jet along with the clustering tree will be passed into RecNN embedding as described 
in the previous subsection.
And the data is generally formatted in hd5f using 
interfaces in deepjets \cite{Barnard:2016qma}.
The details concerning processes used for sample signal and background jets, and other parameters are 
described in the next section.

\noindent\textbf{Preprocessing:}

Before being fed into the RecNN embedding, some basic preprocessing steps are 
applied to the jet samples.
We only apply the necessary translation and rotation to extract ''pure`` jets disconnected to the rest of 
the event. 
\begin{itemize}
 \item translation: jets are translated to be centered at $(\eta, \phi) = 
(0, 0)$.

 \item rotation: the jet is rotated in the ($\eta, \phi$) plane such that the 
energy deposition axis is the 
same (here is (1,0)) for all samples in order to eliminate the effects of global 
influence from
color connections with the remnants of the event (but it will be very 
interesting that we have a solution on event level which takes the color 
connections seriously). Thus only the intra-jet correlation is used for analysis.
The rotation reads as:
\begin{eqnarray}
 \eta &= \hat \eta \cos \alpha + \hat \phi \sin \alpha \\
 \phi &= - \hat \eta \sin \alpha + \hat \phi \cos \alpha
\end{eqnarray}
where the rotation angle $\alpha$ is defined by the "principal axis'' \cite{Almeida:2015jua}:
 \begin{equation}
 \tan \alpha = \frac{\sum_i \frac{\hat \phi_i E_i}{\Delta \hat R_i}}{\sum_i 
\frac{\hat \eta_i E_i}{\Delta \hat R_i}} 
 \end{equation} 
\end{itemize}

\noindent\textbf{Neural Networks Setup:}

The training is carried out with library Scikit-learn \cite{scikit-learn} and 
the RecNN framework\footnote{https://github.com/glouppe/recnn} built by \cite{Louppe:2017ipp}. 
And the subsequent classifier we are using consists of two hidden layers with rectified linear unit 
(ReLU) activations, and the final output node is equiped with sigmoid 
activation. And the Adam \cite{DBLP:journals/corr/KingmaB14} 
algorithm is employed for the minimization.

The input feature vector finally fed into the networks is
${\bf x}_i=(p_i, \eta_i, \phi_i, E_i, E_i/E_J, p_{Ti}, \theta_i = 2 \arctan(\exp(-\eta_i)))$.
The dimension of embedding space is set to be 40 (which has been proven to be large
enough for the case we are examining and also not too large to preserve computation 
time).
The training is done through 10 epoches with the batch size of 64. The learning 
rate is set to be 0.0005 initially and is decayed linearly by a 
rate of 0.9 for every epoch. The training set consists of 50,000 - 100,000 (the exact number
depends on the case) data samples 
and among which 5,000 is used 
as validation set to prevent overfitting. And the performance is tested with a 
dataset of 10,000 - 20,000 samples.

\section{\label{sec:rst}Results}

In this section we show the performance of RecNN on the task of discriminating 
quark/gluon (q/g) jets. And some effects from variants are discussed.
As a byproduct, we also explored the first step on (light-)jet flavor 
identification in Subsection \ref{subsec:ud}.

We show here the discrimination power in use of Receiver Operating 
Characteristic (ROC) curve, which is the standard measure (1/(false positive 
rate) v.s true positive rate) and sometimes is plotted as the signal efficiency
v.s. background rejection rate (i.e., 1 - false positive rate) for a more intuitive presentation.
The Area Under the Curve (AUC) of ROC is the overall 
metric to measure the performance. Generally, the larger the AUC, the better 
the performance.  And as baseline, we also give the 
Boosted Decision Tree (BDT) results using expert features as input.

\subsection{\label{subsec:qg}Quark/Gluon Discrimination}

The processes used to sample jets are $q \bar q\to gg$ and $gg \to gg$ for gluon 
jets, and $gg \to q \bar q$, 
$q\bar q \to q \bar q, qq \to qq$ for light quarks (u,d,s), at $\sqrt{s} = 13$ 
TeV. A set of benchmark jet $p_T$s are examined. The $p_T$ bins are set to be:
[90,110], [180, 220], [480, 520],  [950, 1050] (in GeV).
Jets are clustered using anti-kt algorithm \cite{Cacciari:2008gp} and cone size is set to be $R=0.7$ for
high $p_T$ ($p_T = 1$ TeV) and $R=0.4$ for other relatively lower $p_T$s. 

The performace for different simulation levels are explored: pythia level, delphes e-flow 
and delphes towers. According to our experiments,
using only towers can't provide significant discrimination for 
q/g, thus we don't put the results here.
For pythia level analysis, we employ the constraints of $|\eta|<2.5$,
and discard neutrinos before clustering.

We first investigated the RecNN performance with the architecture indicated in 
the previous section, ROCs are shown and comparison with image approach (CNN) 
and BDT is also carried out. The baseline BDT is composed of scaled mass 
($m_J/p_T$), 
charged particle multiplicity, and girth ($g=\sum_{i \in 
{\rm Jet}} \frac{p_T^i}{p_T^J} r_i$).
For adding information of particle flow, three
scenarios are considered: 
1) without extra particle flow identification; 2) with one-hot vector implementation of 
particle flow identification; 3) with recursively defined pt-weighted charge $Q^{\rm rec}$
instead.
Then the jet-$p_T$ dependence is studied. Furthermore, we 
explored the behavior of several variants: changing cone size for jet clustering;
modification of the input feature set; modification of the embedding feed; etc.

To show the physical implications more clearly, the acceptance and rejection 
rate can be translated into a significance improvement (SI) factor for any 
working point as:
\begin{equation}
 \sigma \equiv \frac{S}{\sqrt{B}} ~ \to ~ \frac{\epsilon_S S}{\sqrt{\epsilon_B 
B}} = \left( \frac{\epsilon_S}{\sqrt{\epsilon_B}} \right) \sigma 
 ~ \to ~ {\rm SI} = \frac{\epsilon_S}{\sqrt{\epsilon_B}}
\end{equation}
Thus, the ROCs can be mapped into significance improvement curves (SICs) 
\cite{Gallicchio:2010dq}. We also show the corresponding SICs in the following.

\begin{table*}[ht!]
\resizebox{1.0\textwidth}{!}{\begin{minipage}{\textwidth}
\begin{center}
\begin{tabular}{|c|c|c|c|} \hline
   & Gluon Jet Efficiency (\%) at 50 \% Quark Jet Acceptance 	& 200 GeV 	& 1000 GeV \\ \hline
  \multirow{4}{*}{Pythia}& BDT of all jet variables			& \textit{5.2}$^*$ &\textit{5.2}$^*$	\\
  & Deep CNN without Color					& 
\textit{4.8}$^*$		& \textit{4.0}$^*$	\\
  & Deep CNN with Color						& 
\textit{4.6}$^*$		& \textit{3.4}$^*$   \\ 
  & \textbf{RecNN} without pflow		& 6.4 & 4.5 \\
\hline \hline
  \multirow{4}{*}{Delphes}& BDT				& 9.5	& 6.2 \\
  & \textbf{RecNN} without pflow		& 7.8	& 4.6\\
  & \textbf{RecNN} with categorical pflow	& 7.1	& 4.5 \\ 
  & \textbf{RecNN} with pt-weighted charge	& 7.8	& 4.9 \\ \hline \hline
  Full Sim.  & DNN@CMS 			& $\sim$\textit{10.0}$^\dagger$	& --\\ \hline
 \end{tabular}
\caption{\label{table}Results displayed for different scenarios for 
comparison. Data with $^*$ is CNN results taken from Ref. 
\cite{Komiske:2016rsd}, and data with $^\dagger$ is take from \cite{CMS-DP-2017-027} (with extrapolation) for quick comparison. 
In the upper part, we show the particle 
level (pythia level) results; 
and in the lower part, detector simulation is included, thus generally reduces 
the discrimination power a bit.}
\end{center}
\end{minipage}}
\end{table*}

In Table \ref{table}, we present the gluon efficiency at the working point of 
50\% quark acceptance, at benchmarks $p_T =200, 1000$ GeV.
We first show the results for pythia level 
in the upper part of the table. And the performance after detector 
simulation is shown in the lower part. 

In Pythia-level, we also take numbers from Ref. \cite{Komiske:2016rsd} for CNNs (pythia level)
as a quick comparison. From the numbers, RecNN is not working much better, but still matches the previous results 
from BDT and CNN. 
In Delphes-level, the RecNN is still giving excellent performance after fast detector simulation.
And compared to pythia level, the delphes level got more influence from $p_T$. The decrease in jet $p_T$  
reduces the performance much quicker. The detector responses at different $p_T$s
deserve more careful investigation.
The numbers show that RecNNs are obviously surpassing BDT and there is  
a potential for full detector simulation. In Ref. \cite{CMS-DP-2017-027}, CMS Collaboration 
has carried out full simulation in DNNs for q/g tagging (CMS Collaboration 
has done the simulation within three different DNN architectures: DeepJet,
LSTM, CNN). We put the number from there (with extrapolation. Since the three DNN architectures give similar 
results for the relevant benchmark working point, we only show one representative number for them.) in Table \ref{table} as reference.

For CNNs within image recognition approach, the particle flow information implemented 
in colored channels \cite{Komiske:2016rsd} (where the detector effects are not 
taken into account) gives slight improvement for 
high $p_T$ jets (500 GeV - 1 TeV) (for $p_T = 1$ TeV, the gluon efficiency 
drops from 4.0\% to 3.4\% for working point of 50\% quark acceptance, and the 
best SI increases by a factor of 1.2), while leaving performance on low 
$p_T$ jets not much changed. 
Since the particle flow is actually a concept closely related to detector
structure, we only consider the implementation of particle flow after 
fast simulation using Delphes.
With extra particle flow information implemented in either one-hot vectors
or recursively defined pt-weighted jet charge, the 
performance is not improved much (although there is a slight increase. And in contrast to CNN,
here the increase is more visible for relatively lower jet $p_T$s.
See Table. \ref{table_auc} for detailed numbers), as claimed in 
\cite{Louppe:2017ipp} for W tagging. This might indicate that by taking particle 
four-momenta directly as input, RecNN is already fully extracting information 
for q/g tagging. Actually the investigation on input features in later part of this section
will also confirm that there is information saturation.


\begin{figure}[ht]
 \begin{center}
  \includegraphics[width=0.45\textwidth]{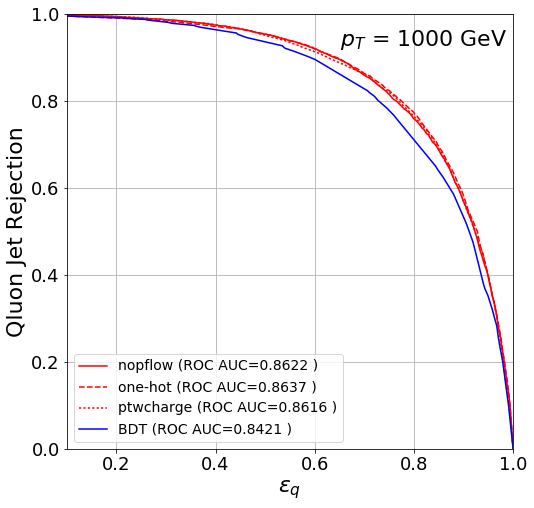}
  \includegraphics[width=0.45\textwidth]{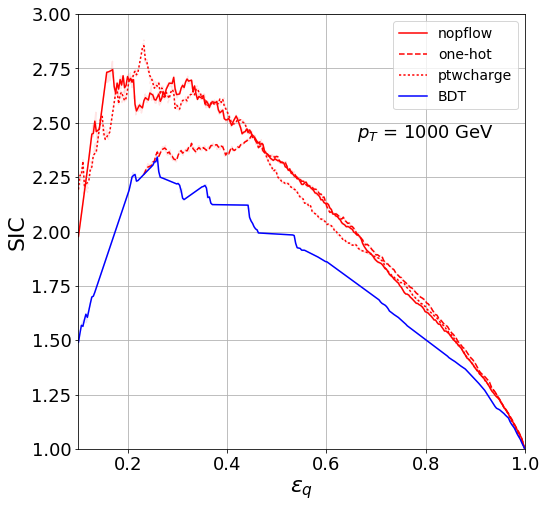}
 \caption{\label{rocs_1tev}ROCs (left) and SICs (right) for jet $p_T = 1$ TeV. 
Baseline BDT and three scenarios concerning particle flow information are 
displayed.``nopflow": no extra particle flow identification is added to RecNN;
``one-hot'': one-hot implementation of particle flow; ``ptwcharge'': recursively
defined pt-weighted charge implemented in the embedding process.}
 \end{center}
\end{figure}

In Fig. \ref{rocs_1tev}, the ROCs and corresponding SICs for 
benchmark transverse momentum $p_T = 1$ TeV are displayed. On the left panel, we show 
the ROCs curves for RecNNs 
and the baseline BDT.
For a signal efficiency of 50\%, the $\sim 95\%$ background can be rejected.
The three RecNNs give better performance than BDT.
And for a signal efficiency of 80\%, the mis-identification rate is about 23\%. 
On the right hand side of Fig. \ref{rocs_1tev}, we show the SICs. The 
significance can be improved by a factor of $\sim$2.75 (best SI) for a signal 
efficiency of 0.2 - 0.3, and by a factor of $\sim$1.7 for $\epsilon_S = 0.8$. 

\begin{figure*}[ht]
 \begin{center}
  \includegraphics[width=0.7\linewidth]{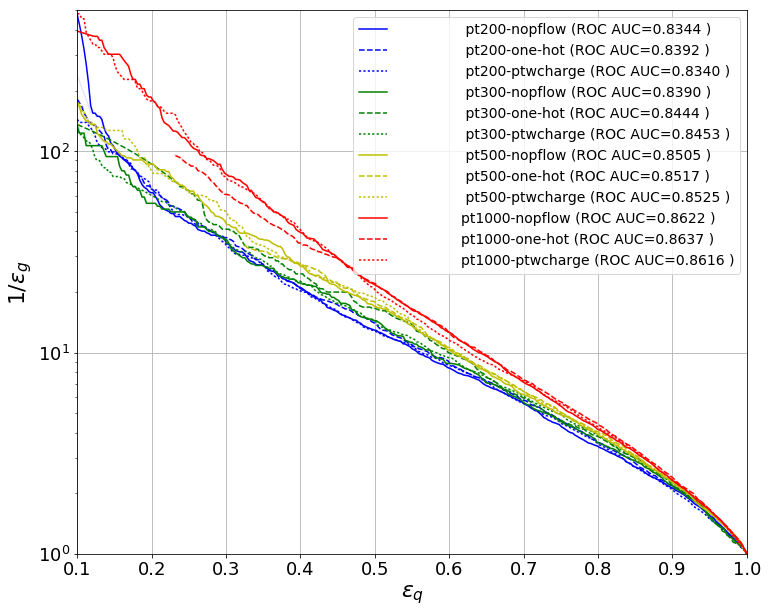}
 \caption{\label{rocs}ROCs of RecNNs for various jet $p_T$s: 200 GeV, 300 GeV, 500 GeV, and 1 TeV.
 Notations are the same as in Fig. \ref{rocs_1tev}.}
 \end{center}
\end{figure*}

\begin{figure*}[ht]
 \begin{center}
   \includegraphics[width=0.7\linewidth]{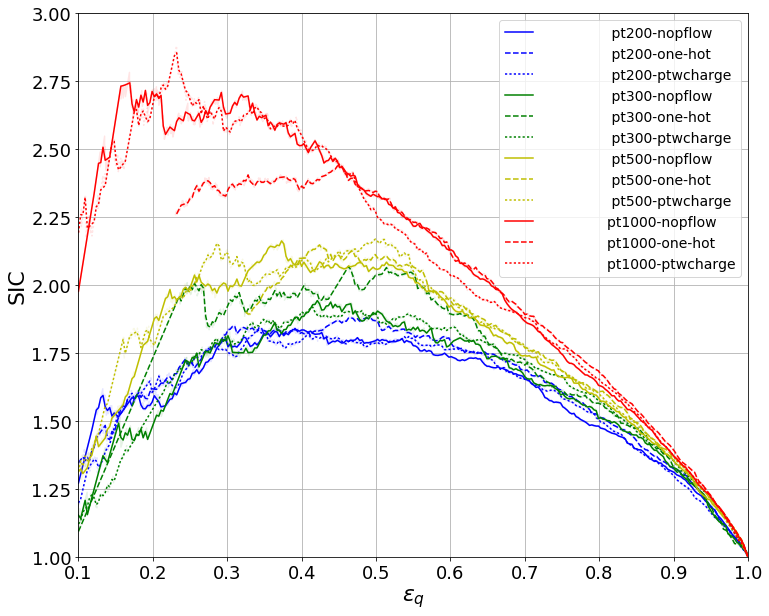}
 \caption{\label{sics}SICs of RecNNs for various jet $p_T$s: 200 GeV, 300 GeV, 500 GeV, and 1 TeV.
  Notations are the same as in Fig. \ref{rocs_1tev}.}
 \end{center}
\end{figure*}

To see the $p_T$ dependence, we show the ROC curves of RecNN 
for different bins of jet $p_T$ in Fig. \ref{rocs}.
Different from Fig. \ref{rocs_1tev}, here we are showing the (1/$\epsilon_g$ v.s $\epsilon_q$)
for a clear comparison. The discriminating power 
increases along with jet $p_T$, which coincides with the behavior in 
conventional approach, since for higher $p_T$, the multiplicity ratio between 
gluon and quark $\frac{N_g}{N_q} \sim \frac{C_A}{C_F} = \frac{9}{4}$ increases 
and slowly reaches its asymptotic limit. The SICs are shown in Fig. \ref{sics}.

For a detailed tabulation of results in numbers, one can find 
the AUCs and background rejection rates for different $p_T$s in Table \ref{table_auc}.
Here the rejection rate is defined as $R_{\epsilon_S} = 1/\epsilon_B 
~@~ \epsilon_S$. 

\begin{table*}[ht!]
\begin{center}
 \begin{tabular}{|c|c|c|c|c|c|c} \hline
  ROC AUC $\mid$ $R_{\epsilon = 80 \%}$  $\mid$ $R_{\epsilon = 50 \%}$  & 200 GeV & 300 GeV & 500 GeV & 1000 GeV \\ \hline
    BDT 				&0.8164 $\mid$ 3.1 $\mid$ 10.5 &0.8443 $\mid$ 3.8 $\mid$ 16.5  &0.8385 $\mid$ 3.5 $\mid$ 14.1 & 0.8421 $\mid$ 3.6 $\mid$ 16.1\\ \hline
  RecNN without pflow identification	&0.8344 $\mid$ 3.4 $\mid$ 12.9& 0.8390 $\mid$ 3.6 $\mid$ 14.4&0.8505 $\mid$ 3.9 $\mid$ 16.9 & 0.8623 $\mid$ 4.2 $\mid$ 21.9  \\ \hline
  RecNN with categorical pflow		&{\bf 0.8392} $\mid$ 3.6 $\mid$ {\bf 14.0}&0.8443 $\mid$ 3.8 $\mid$ {\bf 16.5} &0.8517 $\mid$ 4.0 $\mid$ 17.8& {\bf 0.8637} $\mid$ 4.4 $\mid$ {\bf 22.0} \\ \hline
  RecNN with pt-weighted charge		&0.8340 $\mid$ 3.5 $\mid$ 12.8 &{\bf 0.8453} $\mid$ 3.9 $\mid$ 14.5 &{\bf 0.8525} $\mid$ 4.0 $\mid$ {\bf 18.6}& 0.8616 $\mid$ 4.3 $\mid$ 20.4  \\ \hline
 \end{tabular}
 \end{center}
 \caption{\label{table_auc}AUCs and background rejection rates for different jet $p_Ts$. The baseline BDT and three scenarios 
 concerning particle flow identification are considered. The largest AUCs and $R_{\epsilon=50\%}$s are highlighted in bold.}
\end{table*}

\noindent\textbf{Variants in Network Details and Jet Clustering}

In order to find out how the performance is affected by relevant factors,
several variants
of the procedure are examined.
For simplicity, the experiments are only carried
with samples of $p_T = 200$ GeV. The results are shown in Table. \ref{variants}.
\begin{itemize}
 \item Cone size: larger cone size of jet clustering of 0.7 is examined.
 \item Dismissing the self-representation $u_k$ for embedding (since all the information
 is already contained by the children) in Eqn. \ref{eqn_embedding}, i.e. Eqn. \ref{eqn_embedding1}.
 Thus the number of embedding parameters is reduced almost by 1/3 with
 $W_h \in \mathcal{R}^{q*3q} \to W_h \in 
\mathcal{R}^{q*2q}$. Not much difference is found for the performance,
which means one can even reduce the size of the model
while maintaining the performance.
\begin{equation}
 \textbf{h}_k = \sigma\left( W_h  \left[
    \begin{array}{c}
    \textbf{h}^{\rm jet}_{k_L} \\ \textbf{h}^{\rm jet}_{k_R} 
    \end{array} \right]
    + b_h
  \right) \label{eqn_embedding1}  
\end{equation}
 \item Different input feature sets are tested. The default 
input feature vector is ${\bf x}_i=(p_i, \eta_i, \phi_i, E_i, E_i/E_J, p_{Ti},
\theta_i = 2 \arctan(\exp(-\eta_i)))$. And here other possibilities are experimented: ($p_T$, $\eta$, $\phi$);
($\eta$, $\phi$); ($p_T$); and only particle flow identification (either one-hot implementation (``only one-hot'')
 or recursively defined pt-weighted charge). 
One can find in Table \ref{variants} that for all the sets we have tried, $R_{\epsilon=50\%}$ can
at least reach 11.3 (i.e. mis-identification rate is 8.8\%).
What is intriguing, only particle flow information without any other information from the momentum or 
angular location already gives us a fairly good result. This again corresponds to the dominance of
constituents count. There might be redundant information in the full set.
 \item Multiparton Interaction (MPI): we examined the effects of MPI, and no 
 significant difference was observed.
\end{itemize}

\begin{table}[ht!]
\begin{center}
 \begin{tabular}{c|c|c}\hline 
 Variants & AUC & $R_{\epsilon = 50\%}$\\ \hline \hline
 Baseline		 	& 0.8344 & 12.9 \\ \hline
 R=0.7				& 0.8210 & 12.4 \\ \hline
 $W_h \to R^{q\times2q}$	& 0.8268 & 12.3\\ 
  $W_h \to R^{q\times2q}$ with one-hot	& 0.8313 & 13.7\\ \hline
 \textbf{x}=($p_T$, $\eta$, $\phi$) 	& 0.8291 & 11.8 \\
 \textbf{x}=($\eta$, $\phi$)		& 0.8249 & 11.9 \\
 \textbf{x}=($p_T$)			& 0.8264 & 11.6 \\
 only one-hot				& 0.8255 & 11.9 \\
 \textbf{x}=($Q^{\rm rec}_{\kappa=50\%}$)& 0.8234 & 11.3\\
 \hline
 \end{tabular}
 \caption{\label{variants}Comparison of variants in the network settings. Here ''Baseline''
 is the original RecNN without pflow in Table. \ref{table_auc}.}
\end{center}
\end{table}

These experiments have shown that in the case of q/g discrimination,
RecNN is quite robust against the variances in input features. And there is
still large space for even simplifying the model.
This is partially due to the fact that in q/g tagging the discrete particle count
already dominates, thus most of the information is already contained in the tree
structure itself.

\subsection{\label{subsec:ud}(Light) Quark Jet Flavor}

\begin{figure}[ht!]
 \begin{center}
  \includegraphics[width=0.5\textwidth]{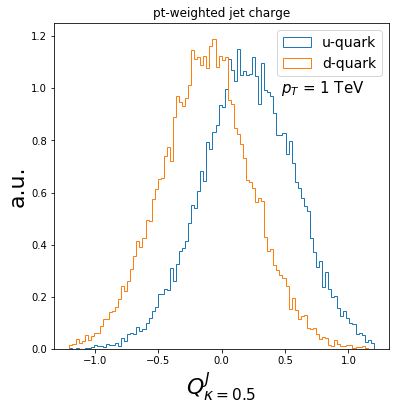}
  \caption{\label{charge_ud}Distributions of pt-weighted jet charge ($\kappa=0.5$) for 
  jets initiating from u-quark and d-quark with $p_T=1$ TeV.}
 \end{center}
\end{figure}

As a bonus, we also checked the RecNN performance on a more difficult task: light quark flavor identification.
Since for all the light quarks, the dominant QCD effects are universal.
And also no heavy-flavor final states, which can have long enough life time such
that leave secondary vertices at the detector, are present.
A possible discriminating element would be remnant electric charge, although the main effects might be
washed out by parton show and hadronization. However it's been shown \cite{Krohn:2012fg}
that even at the LHC, there is still a great potential for measuring jet electric charge.
In Fig. \ref{charge_ud}, we show the distributions of pt-weighted jet charge ($\kappa=0.5$) for u-quark initiating 
jets and d-quark initiating jets with $p_T=1$ TeV.

Jet charge as a useful tool to 
identify initiating light quark flavor has been 
measured at the LHC \cite{Sirunyan:2017tyr,Aad:2015cua}. And it's obviously promising if
we can proceed further in this direction. 

We take u/d discrimination as an attempt.
Jets corresponding to u and d partons are sampled from the production processes of scaled-up W and Z bosons,
at the 13 TeV LHC: $pp \to W'/Z' \to q\bar q$.
Reconstructed jets are matched to the initiating partons by the criterion of 
$\Delta R < 0.5$ between the center of jet and the parton.
The network architecture and the training settings are similar to the ones
in Subsection \ref{subsec:qg}.

\begin{figure}[ht]
 \begin{center}
  \includegraphics[width=0.45\textwidth]{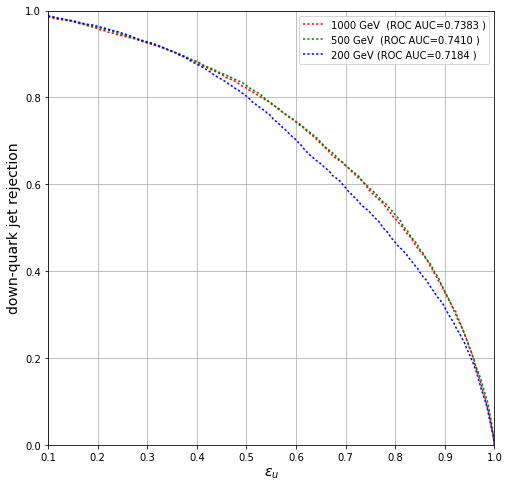}
  \includegraphics[width=0.45\textwidth]{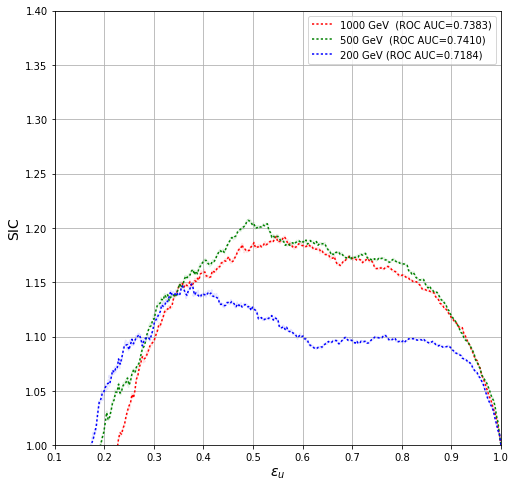}
 \caption{\label{rocs_charge} On the left panel, ROCs of RecNNs are shown for u/d 
discrimination; on the right hand side, corresponding SICs are shown. Benchmark
 $p_T$s include 200 GeV, 500 GeV and 1 TeV.}
 \end{center}
\end{figure}

Results are shown in Fig. \ref{rocs_charge} and Table \ref{ud}. 
The RecNN without particle flow and RecNN with the one-hot implementation are not showing any discriminating power.
Only the hybrid version with pt-weighted charge carried in the embedding is behaving.
We also examined several variants: using only tracks; using tracks and photons. 
The results showed that using only tracks is enough.
A comparison is made between the single observable pt-weighted jet charge and RecNN. 
This hybrid implementation gives matching performance to the pt-weighted charge.

This small experiment shows although no significant improvements
respect to the original observable of jet pt-weighted charge, but it can be
useful for a larger framework for multi-class classification in jet flavors (gluon, b-tagging, 
c-tagging, light flavor identification).
And it also might be useful for boosted W/Z jet tagging.

\begin{table*}[ht!]
\begin{center}
  \begin{tabular}{|c|c|c|c|} \hline
  ROC AUC $\mid$ $R_\epsilon = 50 \%$ 	& 200 GeV & 500 GeV & 1000 GeV 
\\ \hline
  $Q_{\kappa=0.5}$	& 0.7317 $\mid$ 5.3 & 0.7487 $\mid$ 5.9 & 
0.7437 $\mid$ 5.7\\ \hline
  RecNN with pt-weighted charge		&0.7184 $\mid$ 5.1 &0.7410 
$\mid$ 5.8  &0.7383 $\mid$ 5.6 \\ \hline 
 \end{tabular}
 \caption{\label{ud}Discriminating u/d quark with pt-weighted charge information implemented
 in RecNN. The traditional single pt-weighted charge ($Q_{\kappa=0.5}$) is used as baseline.}
\end{center}
\end{table*}

\section{\label{sec:concls}Conclusions}


In Ref. \cite{Louppe:2017ipp}, the framework of RecNNs for jet physics
is built and for the first time tested 
on W/QCD tagging. Motivated by the advantages of efficiency and nice structure of RecNN, and the 
previous findings on its capability, we in this work try to further
explore its application in an advanced topic for jet physics at the LHC: quark/gluon 
discrimination.
And in order to make use of all the 
information collected at the detectors, we also explored the implementation 
of particle flow in RecNN. Besides, some variants for the network details are examined in Section 
\ref{subsec:qg} to 
figure out the relevant factors.
Moreover, a first attempt to identify jet electric charge was made in Section \ref{subsec:ud}
, using recursively embedded pt-weighted charge.
These results show a great potential for RecNNs in broader application and in realistic use.

In the RecNN approach, the raw data taken from detectors are fed into the 
networks directly without information loss.
By embedding the jet data through a tree-like recursive 
process and subsequently feeding into following classifier, 
one has a natural and efficient DNN strategy for jet tagging.
We first investigated the performance of RecNN on pythia-level data, i.e. the 
hadrons without detector effects. 
Then we employed fast detector simulation and took the detector measurements as 
input. 
At this stage, RecNNs
can give discrimination power of rejecting $\sim$95\% gluon jets
at $\epsilon_q=50\%$ for $p_T=1$ TeV.
As for extra particle flow identification implemented in RecNNs,
slight increase can be generally observed (especially for lower jet $p_T$s), but not significant enough.
We examined several variants on the details of the procedure, and interestingly
the results showed that even only with particle flow identification, RecNNs still
give fairly good performance. This might indicate that most of the information 
for q/g discrimination is already contained in the tree-structure itself.
Pile up effects are not taken into account in this work, and jet grooming is also not examined here.
These can be left for future work.

As a byproduct, we also apply the RecNN with recursively defined charge to a more 
difficult task in jet physics: jet flavor (light quarks) identification. It 
actually is the simplest extension from the conventional pt-weighted jet charge 
to its DNN version. And it's showing no better performance, but still gives the 
discrimination power at the same level. We hope it will help in further study on
multi-class classification in jet physics.

Thus as conclusions, we have:
\begin{itemize}
 \item The results with detector simulation indicate a great potential for RecNN in q/g discrimination.
 \item RecNN is robust against the variances in input feature sets. 
 The tree structure itself already contains most of the information in 
q/g discrimination. This is partly due to the fact that the particle multiplicity dominates in q/g tagging.
 \item Extra particle flow identification is not showing significant effects in q/g discrimination, 
 indicating the saturation of input information here.
\end{itemize}

\section{\label{sec:disc}Discussion and Outlook}

There are several interesting aspects and extensions which deserve further investigation.
Here we briefly discuss on a few of them.

\noindent\textbf{Event Level Analysis}

Actually a jet can't be isolated from the remnants of the event, although we 
can get quite ``pure'' jets by grooming. Color connections can be very useful
in many cases.
How to manifest these effects also deserves attention.
And it might be potential if we can deal with them better in event-level analysis.

Easy to be extended to event-level analysis is an important motivation for RecNN.
It is natural to be augmented into lager hierarchical structure. The 
event analysis with only jets has been explored in \cite{Louppe:2017ipp}. A simple RNN 
chain is used there for constructing events from jets.
About the implementation in event-level, how to structure the whole 
event is not trivial.
Every event can be seen as a structured data 
tree, and the whole information of one event just reside within the properties 
of the nodes and links between these nodes. How to properly represent every object
and its connections with other parts of the event would be crucial for designing
NN architectures.

\noindent\textbf{Jet Algorithms as Unsupervised Learning Procedure}

In the framework of DNN, 
accordingly adjusting the jet clustering might help to gain better performance.
Actually
jet finding itself can be treated as a minimization problem
\cite{Angelini:2004ac,Grigoriev:2003yc,Grigoriev:2003tn,Volobouev:2009rv},
thus it will be very interesting if we can  
naturally include the process of jet finding in event-level implementation.

\noindent\textbf{Application: New Physics}

Generally speaking, new physics will have interesting patterns when concerned with their new 
particle spectrum and decay modes. For example, SUSY 
events will generally produce large amounts of final states (more complicated 
hierarchical structure), and sometimes several soft leptons in electroweakino 
search. Whether DNNs have better tolerance on this kind of topologies is also worthwhile
for investigation.

\begin{acknowledgements}
This work is partially supported by Natural Science Foundation of China under
grant no. 11475180.
And the author would like to thank Lei Wang for discussion on ML application 
in condensed matter physics, and giving some nice advice; Zhao Li and Hao Zhang 
for collaboration which inspires part of this work.
More importantly, a big thank-you goes to Gilles Louppe 
for information on some techincal details of the RecNN package and some nice discussions,
and Edmund Noel Dawe for updating the python version of the deepjets package. 
The author would also like to acknowledge the MCnet 
summer school during which part of this work was carried out.
\end{acknowledgements}

\bibliography{ref}

\begin{thebibliography}{10}

\bibitem{Larkoski:2017jix}
Andrew~J. Larkoski, Ian Moult, and Benjamin Nachman.
\newblock {Jet Substructure at the Large Hadron Collider: A Review of Recent
  Advances in Theory and Machine Learning}.
\newblock 2017, 1709.04464.

\bibitem{Adams:2015hiv}
D.~Adams et~al.
\newblock {Towards an Understanding of the Correlations in Jet Substructure}.
\newblock {\em Eur. Phys. J.}, C75(9):409, 2015, 1504.00679.

\bibitem{Cogan:2014oua}
Josh Cogan, Michael Kagan, Emanuel Strauss, and Ariel Schwarztman.
\newblock {Jet-Images: Computer Vision Inspired Techniques for Jet Tagging}.
\newblock {\em JHEP}, 02:118, 2015, 1407.5675.

\bibitem{Almeida:2015jua}
Leandro~G. Almeida, Mihailo Backović, Mathieu Cliche, Seung~J. Lee, and Maxim
  Perelstein.
\newblock {Playing Tag with ANN: Boosted Top Identification with Pattern
  Recognition}.
\newblock {\em JHEP}, 07:086, 2015, 1501.05968.

\bibitem{Pearkes:2017hku}
Jannicke Pearkes, Wojciech Fedorko, Alison Lister, and Colin Gay.
\newblock {Jet Constituents for Deep Neural Network Based Top Quark Tagging}.
\newblock 2017, 1704.02124.

\bibitem{Kasieczka:2017nvn}
Gregor Kasieczka, Tilman Plehn, Michael Russell, and Torben Schell.
\newblock {Deep-learning Top Taggers or The End of QCD?}
\newblock {\em JHEP}, 05:006, 2017, 1701.08784.

\bibitem{Baldi:2016fql}
Pierre Baldi, Kevin Bauer, Clara Eng, Peter Sadowski, and Daniel Whiteson.
\newblock {Jet Substructure Classification in High-Energy Physics with Deep
  Neural Networks}.
\newblock {\em Phys. Rev.}, D93(9):094034, 2016, 1603.09349.

\bibitem{Guest:2016iqz}
Daniel Guest, Julian Collado, Pierre Baldi, Shih-Chieh Hsu, Gregor Urban, and
  Daniel Whiteson.
\newblock {Jet Flavor Classification in High-Energy Physics with Deep Neural
  Networks}.
\newblock {\em Phys. Rev.}, D94(11):112002, 2016, 1607.08633.

\bibitem{Barnard:2016qma}
James Barnard, Edmund~Noel Dawe, Matthew~J. Dolan, and Nina Rajcic.
\newblock {Parton Shower Uncertainties in Jet Substructure Analyses with Deep
  Neural Networks}.
\newblock {\em Phys. Rev.}, D95(1):014018, 2017, 1609.00607.

\bibitem{deOliveira:2015xxd}
Luke de~Oliveira, Michael Kagan, Lester Mackey, Benjamin Nachman, and Ariel
  Schwartzman.
\newblock {Jet-images — deep learning edition}.
\newblock {\em JHEP}, 07:069, 2016, 1511.05190.

\bibitem{Komiske:2016rsd}
Patrick~T. Komiske, Eric~M. Metodiev, and Matthew~D. Schwartz.
\newblock {Deep learning in color: towards automated quark/gluon jet
  discrimination}.
\newblock {\em JHEP}, 01:110, 2017, 1612.01551.

\bibitem{CMS-DP-2017-005}
CMS Collaboration.
\newblock {Heavy flavor identification at CMS with deep neural networks}.
\newblock Mar 2017.
\newblock \url{https://cds.cern.ch/record/2255736}.

\bibitem{Goodfellow-et-al-2016}
Ian Goodfellow, Yoshua Bengio, and Aaron Courville.
\newblock {\em Deep Learning}.
\newblock MIT Press, 2016.
\newblock \url{http://www.deeplearningbook.org}.

\bibitem{Louppe:2017ipp}
Gilles Louppe, Kyunghyun Cho, Cyril Becot, and Kyle Cranmer.
\newblock {QCD-Aware Recursive Neural Networks for Jet Physics}.
\newblock 2017, 1702.00748.

\bibitem{Aad:2016oit}
Georges Aad et~al.
\newblock {Measurement of the charged-particle multiplicity inside jets from
  $\sqrt{s}=8$ TeV $pp$ collisions with the ATLAS detector}.
\newblock {\em Eur. Phys. J.}, C76(6):322, 2016, 1602.00988.

\bibitem{Gallicchio:2012ez}
{Quark and Gluon Jet Substructure}.
\newblock {\em JHEP}, 04:090, 2013, 1211.7038.

\bibitem{ATL-PHYS-PUB-2017-017}
ATLAS Collaboration.
\newblock {Quark versus Gluon Jet Tagging Using Jet Images with the ATLAS
  Detector}.
\newblock Technical Report ATL-PHYS-PUB-2017-017, CERN, Geneva, Jul 2017.

\bibitem{CMS-DP-2017-027}
CMS Collaboration.
\newblock {New Developments for Jet Substructure Reconstruction in CMS}.
\newblock Jul 2017.
\newblock \url{https://cds.cern.ch/record/2275226}.

\bibitem{Gras:2017jty}
Philippe Gras, Stefan Höche, Deepak Kar, Andrew Larkoski, Leif Lönnblad,
  Simon Plätzer, Andrzej Siódmok, Peter Skands, Gregory Soyez, and Jesse
  Thaler.
\newblock {Systematics of quark/gluon tagging}.
\newblock {\em JHEP}, 07:091, 2017, 1704.03878.

\bibitem{Salam:2009jx}
Gavin~P. Salam.
\newblock {Towards Jetography}.
\newblock {\em Eur. Phys. J.}, C67:637--686, 2010, 0906.1833.

\bibitem{Datta:2017rhs}
Kaustuv Datta and Andrew Larkoski.
\newblock {How Much Information is in a Jet?}
\newblock {\em JHEP}, 06:073, 2017, 1704.08249.

\bibitem{pmlr-v15-glorot11a}
Xavier Glorot, Antoine Bordes, and Yoshua Bengio.
\newblock Deep sparse rectifier neural networks.
\newblock In Geoffrey Gordon, David Dunson, and Miroslav Dudík, editors, {\em
  Proceedings of the Fourteenth International Conference on Artificial
  Intelligence and Statistics}, volume~15 of {\em Proceedings of Machine
  Learning Research}, pages 315--323, Fort Lauderdale, FL, USA, 11--13 Apr
  2011. PMLR.

\bibitem{Sirunyan:2017ulk}
Albert~M Sirunyan et~al.
\newblock {Particle-flow reconstruction and global event description with the
  CMS detector}.
\newblock {\em JINST}, 12(10):P10003, 2017, 1706.04965.

\bibitem{Field:1977fa}
R.~D. Field and R.~P. Feynman.
\newblock {A Parametrization of the Properties of Quark Jets}.
\newblock {\em Nucl. Phys.}, B136:1, 1978.

\bibitem{Krohn:2012fg}
David Krohn, Matthew~D. Schwartz, Tongyan Lin, and Wouter~J. Waalewijn.
\newblock {Jet Charge at the LHC}.
\newblock {\em Phys. Rev. Lett.}, 110(21):212001, 2013, 1209.2421.

\bibitem{Sjostrand:2014zea}
Torbjörn Sjöstrand, Stefan Ask, Jesper~R. Christiansen, Richard Corke,
  Nishita Desai, Philip Ilten, Stephen Mrenna, Stefan Prestel, Christine~O.
  Rasmussen, and Peter~Z. Skands.
\newblock {An Introduction to PYTHIA 8.2}.
\newblock {\em Comput. Phys. Commun.}, 191:159--177, 2015, 1410.3012.

\bibitem{deFavereau:2013fsa}
J.~de~Favereau, C.~Delaere, P.~Demin, A.~Giammanco, V.~Lemaître, A.~Mertens,
  and M.~Selvaggi.
\newblock {DELPHES 3, A modular framework for fast simulation of a generic
  collider experiment}.
\newblock {\em JHEP}, 02:057, 2014, 1307.6346.

\bibitem{Cacciari:2011ma}
Matteo Cacciari, Gavin~P. Salam, and Gregory Soyez.
\newblock {FastJet User Manual}.
\newblock {\em Eur. Phys. J.}, C72:1896, 2012, 1111.6097.

\bibitem{Cacciari:2008gp}
Matteo Cacciari, Gavin~P. Salam, and Gregory Soyez.
\newblock {The Anti-k(t) jet clustering algorithm}.
\newblock {\em JHEP}, 04:063, 2008, 0802.1189.

\bibitem{scikit-learn}
F.~Pedregosa, G.~Varoquaux, A.~Gramfort, V.~Michel, B.~Thirion, O.~Grisel,
  M.~Blondel, P.~Prettenhofer, R.~Weiss, V.~Dubourg, J.~Vanderplas, A.~Passos,
  D.~Cournapeau, M.~Brucher, M.~Perrot, and E.~Duchesnay.
\newblock Scikit-learn: Machine learning in {P}ython.
\newblock {\em Journal of Machine Learning Research}, 12:2825--2830, 2011.

\bibitem{DBLP:journals/corr/KingmaB14}
Diederik~P. Kingma and Jimmy Ba.
\newblock Adam: {A} method for stochastic optimization.
\newblock {\em CoRR}, abs/1412.6980, 2014, 1412.6980.

\bibitem{Gallicchio:2010dq}
Jason Gallicchio, John Huth, Michael Kagan, Matthew~D. Schwartz, Kevin Black,
  and Brock Tweedie.
\newblock {Multivariate discrimination and the Higgs + W/Z search}.
\newblock {\em JHEP}, 04:069, 2011, 1010.3698.

\bibitem{Sirunyan:2017tyr}
Albert~M Sirunyan et~al.
\newblock {Measurements of jet charge with dijet events in pp collisions at
  $\sqrt{s}=8$ TeV}.
\newblock {\em JHEP}, 10:131, 2017, 1706.05868.

\bibitem{Aad:2015cua}
Georges Aad et~al.
\newblock {Measurement of jet charge in dijet events from $\sqrt{s}$=8  TeV
  pp collisions with the ATLAS detector}.
\newblock {\em Phys. Rev.}, D93(5):052003, 2016, 1509.05190.

\bibitem{Angelini:2004ac}
L.~Angelini, G.~Nardulli, L.~Nitti, M.~Pellicoro, D.~Perrino, and
  S.~Stramaglia.
\newblock {Deterministic annealing as a jet clustering algorithm in hadronic
  collisions}.
\newblock {\em Phys. Lett.}, B601:56--63, 2004, hep-ph/0407214.

\bibitem{Grigoriev:2003yc}
D.~{\relax Yu}. Grigoriev, E.~Jankowski, and F.~V. Tkachov.
\newblock {Towards a standard jet definition}.
\newblock {\em Phys. Rev. Lett.}, 91:061801, 2003, hep-ph/0301185.

\bibitem{Grigoriev:2003tn}
D.~{\relax Yu}. Grigoriev, E.~Jankowski, and F.~V. Tkachov.
\newblock {Optimal jet finder}.
\newblock {\em Comput. Phys. Commun.}, 155:42--64, 2003, hep-ph/0301226.

\bibitem{Volobouev:2009rv}
I.~Volobouev.
\newblock {FFTJet: A Package for Multiresolution Particle Jet Reconstruction in
  the Fourier Domain}.
\newblock 2009, 0907.0270.

\end{thebibliography}
\bibliographystyle{hunsrt}

\end{document}